\documentclass[fleqn]{llncs}
\usepackage{pstricks,pst-tree}
\usepackage{amsmath,amssymb}
\title{Towards ``Propagation = Logic + Control''}
\author{Sebastian Brand\inst{1} \and Roland H.C. Yap\inst{2}}
\institute{National ICT Australia, Victoria Research Lab, Melbourne, Australia
\and School of Computing, National University of Singapore, Singapore}

\begin{document}
\maketitle

\begin{abstract}
Constraint propagation algorithms implement logical inference.
For efficiency, it is essential to control whether
and in what order basic inference steps are taken.
We provide a high-level framework that clearly differentiates
between information needed for controlling propagation
versus that needed for the logical semantics of complex
constraints composed from primitive ones.
We argue for the appropriateness of our \emph{controlled propagation} framework
by showing that it captures the underlying principles of
manually designed propagation algorithms,
such as literal watching for unit clause propagation
and the lexicographic ordering constraint.
We provide an implementation and benchmark results that demonstrate
the practicality and efficiency of our framework.
\end{abstract}


\renewcommand{\leq}		{\leqslant}
\renewcommand{\geq}		{\geqslant}
\renewcommand{\nleq}		{\nleqslant}
\renewcommand{\ngeq}		{\ngeqslant}
\renewcommand{\emptyset}	{\varnothing}

\newcommand{\colorctrl}		{}
\newcommand{\colorlogic}	{}

\newcommand{\bfemph}[1]		{\textbf{\emph{#1}}}

\newcommand{\tuple}[1]		{\langle #1 \rangle}
\newcommand{\Tuple}[1]		{\left\langle #1 \right\rangle}

\newcommand{\rangebopd}[5]	{#1_#2#4 #5 #4#1_#3}
\newcommand{\rangeb}[3]		{\rangebopd{#1}{#2}{#3},{\ldots}}
\newcommand{\rangenop}[2]	{\rangebopd{#1}1n{#2}{\ldots}}
\newcommand{\rangen}[1]		{\rangebopd{#1}1n,{\ldots}}

\newcommand{\Eg}		{E.\,g.}
\newcommand{\eg}		{e.\,g.}
\newcommand{\ie}		{i.\,e.}

\newcommand{\eclipse}		{\textup{ECL\textsuperscript{\textit{i}}PS%
				\textsuperscript{\textit{e}}}}

\newcommand{\qtext}[1]		{\quad\text{#1}\quad}

\newcommand{\setintersection}	{\cap}
\newcommand{\setunion}		{\cup}

\newcommand{\lAnd}		{\bigwedge}
\newcommand{\lOr}		{\bigvee}

\newcommand{\limplies}		{\rightarrow}
\newcommand{\limpliesd}		{\Rightarrow}
\newcommand{\qlimpliesd}	{\quad\Rightarrow\quad}
\newcommand{\entails}		{\qlimpliesd}

\newcommand{\constraint}[1]	{\mathsf{#1}}
\newcommand{\clause}		{\constraint{clause}}
\newcommand{\alldifft}		{\constraint{alldifferent\_tp}}
\newcommand{\diff}		{\constraint{different\_tp}}
\newcommand{\lex}		{\constraint{lex}}
\newcommand{\cor}		{\constraint{or}}
\newcommand{\cand}		{\constraint{and}}
\newcommand{\cnot}		{\constraint{not}}

\newcommand{\controlflag}[1]	{\mbox{\colorctrl\textsl{#1}}}
\newcommand{\cf}		{\controlflag{chk-false}}
\newcommand{\ct}		{\controlflag{chk-true}}
\newcommand{\ir}		{\controlflag{irrelevant}}

\newcommand{\del}		{\text{\colorctrl delete}}
\newcommand{\ctrl}		{\mathcal{F}}
\newcommand{\isrelevant}	{\mathit{is\_relevant}}

\newcommand{\setadd}		{\mathbin{{\cup}{=}}}
\newcommand{\trails}		{\rightsquigarrow}

\newcommand{\cw}[2]		{#1 \text{ with } #2}
\newcommand{\cwt}[2]		{#1 \text{ with } \tuple{#2}}
\newcommand{\cbw}[3]		{#1 \equiv #2 \text{ with } #3}
\newcommand{\cb}[2]		{#1 \equiv #2}

\newcommand{\lentailed}		{\vartriangleright}
\newcommand{\impl}		{\lentailed}
\newcommand{\corimpl}		{\constraint{or}_{\impl}}

\newcommand{\gor}		{$\constraint{or}$}
\newcommand{\gand}		{$\constraint{and}$}
\newcommand{\gnot}		{$\constraint{not}$}

\newcommand{\Trf}[1]		{\Tr{\psframebox{#1}}}

\psset{treesep=4ex,levelsep=6ex,framearc=0.7,arrows=->}


\section{Introduction}

Constraint programming solves combinatorial problems
by combining search and logical inference.
The latter, constraint propagation, aims at reducing the search space.
Its applicability and usefulness relies on the availability
of efficiently executable propagation algorithms.

It is well understood how primitive constraints, \eg\ indexical constraints,
and also their reified versions, are best propagated.
We also call such primitive constraints \emph{pre-defined},
because efficient, special-purpose propagation algorithms exist for them
and many constraint solving systems provide implementations.
However, when modelling problems, one often wants to make use
of more complex constraints whose semantics can best be described as
a combination of pre-defined constraints using logical operators
(\ie\ conjunction, disjunction, negation).
Examples are constraints for breaking symmetries \cite{frisch:2002:global}
and channelling constraints \cite{cheng:1999:increasing}.

Complex constraints are beneficial in two aspects.
Firstly, from a reasoning perspective, complex constraints give
a more direct and understandable high-level problem model.
Secondly, from a propagation perspective, the more more global scope
of such constraints can allow stronger inference.
While elaborate special-purpose propagation algorithms are known
for many specific complex constraints
(the classic example is the $\constraint{alldifferent}$ constraint
discussed in \cite{regin:1994:filtering}),
the diversity of combinatorial problems tackled
with constraint programming in practice
implies that more diverse and rich constraint languages are needed.

Complex constraints which are defined by logical combinations of
primitive constraints can always be decomposed into their primitive
constituents and Boolean constraints, for which propagation methods exist.
However, decomposing in this way may
\begin{itemize}
\item[({\sffamily A})\!]\ 	cause redundant propagation, as well as
\item[({\sffamily B})\!]\ 	limit possible propagation.
\end{itemize}
This is due to the loss of a global view:
information between the constituents of a decomposition
is only exchanged via shared constrained variables.

\smallskip
As an example, consider the implication constraint
$x = 5 \limplies y \neq 8$ during constructive search.
First, once the domain of $x$ does not contain $5$ any more,
the conclusion $y \neq 8$ is irrelevant for the remainder of the search.
Second, only an instantiation of $y$ is relevant as
non-instantiating reductions of the domain of $y$ do not allow
any conclusions on $x$.
These properties are lost if
the implication is decomposed into the reified constraints
$(x = 5) \equiv b_1$, $(y \neq 8) \equiv b_2$ and the Boolean constraints
$\cnot(b_1, b_1')$, $\cor(b_1', b_2)$.
\smallskip

Our focus is point ({\sffamily A}).  We show how \bfemph{shared control information}
allows a constraint to signal others what sort of information is relevant
to its propagation or that any future propagation on their part has become
irrelevant to it.
We address ({\sffamily B}) to an extent by considering implied constraints
in the decomposition.
Such constraints may be logically redundant but not operationally so.
Control flags connecting them to their respective antecedents
allow us to keep track of the special status of implied constraint,
so as to avoid redundant propagation steps.
Our proposed control framework is naturally applicable not only
to the usual tree-structure decomposition but also to those with
a more complex DAG structure, which permits stronger propagation.

Our objective is to capture the \emph{essence} of manually designed
propagation algorithms,
which implicitly merge the separate aspects of logic and control.
We summarise this by \emph{Propagation = Logic + Control}
in the spirit of \cite{kowalski:1979:algorithm}.
The ultimate goal of our approach is a fully automated treatment
of arbitrary complex constraints
specified in a logic-based constraint definition language.
We envisage such a language to be analogous to CLP but focused on propagation.
Our framework would allow users
lacking the expertise in or the time for the development
of specialised propagation
to rapidly prototype and refine propagation algorithms for complex constraints.


\subsection*{Preliminaries}

Consider a finite sequence of different variables
$X = \rangeb x1m$ with respective domains
$D(x_1), \ldots, D(x_m)$.
A \bfemph{constraint} $C$ on $X$ is a pair $\tuple{S, X}$.
The set $S$ is an $m$-ary relation and a subset of
the Cartesian product of the domains, that is,
$S \subseteq D(x_1) \times \ldots \times D(x_m)$.
The elements of $S$ are the \bfemph{solutions} of the constraint,
and $m$ is its \emph{arity}.  We assume $m \geq 1$.
We sometimes write $C(X)$ for the constraint
and often identify $C$ with $S$.

We distinguish pre-defined, \bfemph{primitive} constraints,
such as $x = y, x \leq y$, and \bfemph{complex} constraints,
constructed from the primitive constraints and
the logical operators $\lor, \land, \lnot$ etc.
For each logical operator there is a corresponding
\bfemph{Boolean} constraint.
For example, the satisfying assignments of $x \lor y = z$ are
the solutions of the constraint $\cor(x,y,z)$.
The \bfemph{reified} version of a constraint $C(X)$ is
a constraint on $X$ and an additional Boolean variable $b$
reflecting the truth of $C(X)$; we write it as $\cb {C(X)}b$.
Complex constraints can be \bfemph{decomposed} into a set of
reified primitive constraints and Boolean constraints,
whereby new Boolean variables are introduced.
For example, the first step in decomposing $C_1 \lor C_2$ may result
in the three constraints
$\cb {C_1}{b_1}$, $\cb {C_2}{b_2}$, and $\cor(b_1, b_2, 1)$.

Constraint \bfemph{propagation} aims at inferring new constraints from
given constraints.  In its most common form,
a single constraint is considered, and
the domains of its variables are reduced
without eliminating any solution of the constraint.
If every domain is maximally reduced and none is empty,
the constraint is said to be \bfemph{domain-consistent} (DC).
For instance, $x < y$ with $D(x) = \{1,2\}$, $D(y) = \{1,2,3\}$
can be made domain-consistent by inferring the constraint $y \neq 1$,
leading to the smaller domain $D(y) = \{2,3\}$.

Decomposing a complex constraint may hinder propagation.
For example, DC-establishing propagation
is guaranteed to result in the same domain reductions
on a constraint and its decomposition only if the constraint graph
of the decomposition is a tree \cite{freuder:1982:sufficient}.
For instance, the constraints of the decomposition
of the constraint $(x > y) \land (x < y)$ considered in isolation
do not indicate its inconsistency.


\section{Logic and Control Information}

A complex constraint expressed as a logical combination of primitive
constraints can be decomposed into its primitive parts.
However, such a naive decomposition has the disadvantage that it assigns
equal relevance to every constraint.
This may cause redundant reasoning to take place for the individual
primitive constraints and connecting Boolean constraints.
We prevent this by maintaining fine-grained control information
on whether the \emph{truth} or \emph{falsity} of
individual constraints matters.
We say that a truth status of a constraint is \bfemph{relevant}
if it entails the truth status of some other constraint.

We focus on the disjunction operator first.
\begin{proposition}\label{prop:disj}
Suppose $C$ is the disjunctive constraint $C_1 \lor C_2$.
Consider the truth status of $C$ in terms
of the respective truth statuses of the individual
constraints $C_1$, $C_2$.
\begin{itemize}
\item If the falsity of $C$ is asserted then
	the falsity of $C_1$ and $C_2$ can be asserted.
\item If the truth of $C$ is asserted then
	the falsity of $C_1$ and $C_2$ is relevant, but not their truth.
\item If the truth of $C$ is queried then
	the truth of $C_1$ and $C_2$ is relevant, but not their falsity.
\item If the falsity of $C$ is queried then
	the falsity of only one of $C_1$ or $C_2$ is relevant,
	but not the their truth.
\end{itemize}
\end{proposition}
\begin{proof}
Let the reified version of $C$ be $\cb {(C_1 \lor C_2)}b$
and its partial decomposition be
$\cb {C_1}{b_1}$, $\cb {C_2}{b_2}$, $\cor(b_1, b_2, b)$.
The following cases can occur when asserting or querying $C$.
\begin{description}
\item[Case $b=0$.]
	Then $C_1$ and $C_2$ must both be asserted to be false.

\item[Case $b=1$.]\mbox{}
\begin{itemize}
\item Suppose $C_1$ is found to be true.
	This means that both the truth and the falsity of $C_2$,
	hence $C_2$ itself, have become irrelevant for the remainder
	of the current search.
	Although this simplifies the representation of $C$ to $C_1$,
	it does not lead to any inference on it.
	In this sense, the truth of $C_1$ is useless information.

	The case of $C_2$ being true is analogous.
\item Suppose $C_1$ is found to be false.
	This is useful information as we now must assert the truth of $C_2$,
	which may cause further inference in $C_2$.

	The case of $C_2$ being false is analogous.
\end{itemize}
\smallskip
	Only falsity of $C_1$ or $C_2$ is information
	that may cause propagation.
	Their truth is irrelevant in this respect.

\item[Case $b$ unknown.]\mbox{}
We now assume that we know what aspect of the truth status
of $C$ is relevant: its truth or its falsity.
If neither is relevant then we need not consider $C$,
\ie\ $C_1$ and $C_2$, at all.
If both the truth and falsity of $C$ are relevant,
the union of the individual cases applies.
\begin{description}
\item[Truth of $C$ is queried:]\mbox{}
\begin{itemize}
\item Suppose $C_1$ or $C_2$ is found to be true.
	This means that $C$ is true, and knowing
	either case is therefore useful information.
\item Suppose $C_1$ is found to be false.
	Then the truth of $C$ depends on the truth of $C_2$.
	The reasoning for $C_2$ being false is analogous.
\end{itemize}
\smallskip
	The truth of both $C_1$ and $C_2$ matters,
	but not their falsity.
\item[Falsity of $C$ is queried:]\mbox{}
\begin{itemize}
\item Suppose $C_1$ or $C_2$ is found to be true.
	While this means that $C$ is true,
	this is not relevant since its falsity is queried.
\item Suppose $C_1$ is found to be false.
	Then the falsity of $C$ depends on the falsity of $C_2$.
	Now suppose otherwise that $C_1$ is queried for
	falsity but \emph{not} found to be false.
	If $C_1$ is not false then $C$ cannot be false.
	It is important to realise that
	this reasoning is independent of $C_2$.

	The reasoning for $C_2$ being false is symmetric.
\end{itemize}
\smallskip
	In summary, to determine the falsity of $C$, it suffices to
	query the falsity of \emph{just one} of $C_1$ or $C_2$.
\qed
\end{description}
\end{description}
\end{proof}

\begin{figure}[t]
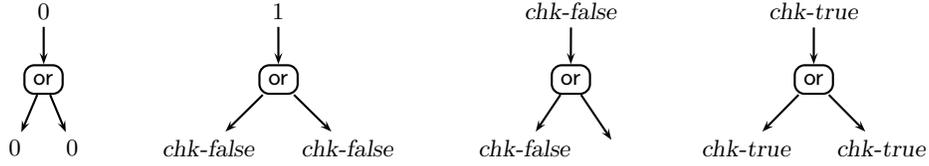

\pstree[nodesepA=3pt]{\Tr{\colorlogic 0}}{\pstree[nodesepA=0pt,nodesepB=3pt]{
	\Trf{\gor}}{%
	\Tr{\colorlogic 0}
	\Tr{\colorlogic 0}
}}
\hfill
\pstree[nodesepA=3pt]{\Tr{\colorlogic 1}}{\pstree[nodesepA=0pt,nodesepB=3pt]{
	\Trf{\gor}}{%
	\Tr{\colorctrl $\cf$}
	\Tr{\colorctrl $\cf$}
}}
\hfill
\pstree[nodesepA=3pt]{\Tr{\colorctrl $\cf$}}{\pstree[nodesepA=0pt,nodesepB=3pt]{
	\Trf{\gor}}{%
	\Tr{\colorctrl $\cf$}
	\Tr{}
}}
\hfill
\pstree[nodesepA=3pt]{\Tr{\colorctrl $\ct$}}{\pstree[nodesepA=0pt,nodesepB=3pt]{
	\Trf{\gor}}{%
	\Tr{\colorctrl $\ct$}
	\Tr{\colorctrl $\ct$}
}}
\caption{Control flow through a disjunction}
\label{fig:control-flow-or}
\end{figure}

Fig.~\ref{fig:control-flow-or}
shows the flow of control information through a disjunction.
There, and throughout the rest of this paper,
we denote a truth query by $\ct$ and a falsity query by $\cf$.

Analogous studies on control flow can be conducted for all other Boolean operators.
The case of a negated constraint is straightforward:
truth and falsity swap their roles.
Conjunction is entirely symmetric to disjunction due to De Morgan's law.
For example, a query for falsity of the
conjunction propagates to both conjuncts while a query for truth
need only be propagated to one conjunct.
We remark that one can apply such an analysis to other kinds of operators
including non-logical ones.
Thus, the $\constraint{cardinality}$ constraint
\cite{hentenryck:1991:cardinality}
can be handled within this framework.


\subsection{Controlled Propagation}

Irrelevant inference can be prevented by distinguishing
whether the truth or the falsity of a constraint matters.
This control information arises from truth information
and is propagated similarly.
By \bfemph{controlled propagation} we mean constraint propagation that
(1) conducts inference according to truth and falsity information and
(2) propagates such information.

We now characterise controlled propagation
for a complex constraint in decomposition.
We are interested in the \emph{effective} propagation, \ie\ newly inferred
constraints (such as smaller domains) on the original variables
rather than on auxiliary Boolean variables.
We assume that only individual constraints are propagated%
\footnote{\Eg, path-consistency enforcing propagation
	considers two constraints at a time.}.
This is the usual case in practice.
\begin{theorem}[Controlled Propagation]
Controlled and uncontrolled propagation of the constraints of the
decomposition of a constraint $C$ are equivalent with respect to
the variables of $C$ if only single constraints are propagated.
\qed
\end{theorem}
\begin{proof}
Proposition~\ref{prop:disj} and analogous propositions
for the other Boolean operators.
\end{proof}

In the following, we explain a formal framework
for maintaining and reacting to control information.


\subsubsection{Control store.}

Constraints communicate truth information by shared Boolean variables.
Similarly, we think of control information being communicated
between constraints by shared sets of control flags.
As control flags we consider the truth status queries $\ct$, $\cf$
and the additional flag $\ir$ signalling \bfemph{permanent irrelevance}.
In this context, `permanently' refers to
subsidiary parts of the search, that is, until the next back-tracking.
Note that the temporary absence of truth and falsity queries on a constraint
is not the same as its irrelevance.
We write
\[
	\cw C{\mathcal{FS}}
\]
to mean that the constraint $C$ can read and update the
sequence of control flag sets $\mathcal{FS}$.
One difference between logic and control information communication
is that control flows only one way, from a producer to a consumer.


\subsubsection{Propagating control.}

A set of control flags $\ctrl$ is updated by adding or deleting flags.
We abbreviate the adding operation $\ctrl := \ctrl \setunion \{\controlflag f\}$
as $\ctrl \setadd \controlflag f$.
We denote by $\ctrl_1 \trails \ctrl_2$
that from now on permanently changes to the control flags in $\ctrl_1$
are reflected in corresponding changes to $\ctrl_2$;
\eg\ an addition of $\controlflag f$ to $\ctrl_1$ leads to an
addition of $\controlflag f$ to $\ctrl_2$.

We employ rules to specify how control information is attached
to the constituents of a decomposed complex constraint,
and how it propagates.
The rule $A \limpliesd B$ denotes that the conditions in $A$,
consisting of constraints and associated control information,
entail the constraints and the updates of
control information specified in $B$.
We use \bfemph{delete} statements in the conclusion
to explicitly remove a constraint from the constraint store
once it is solved or became permanently irrelevant.


\subsubsection{Relevance.}

At the core of controlled propagation is the principle that reasoning effort
should be made only if it is relevant to do so, that is,
if the truth or falsity of the constraint at hand is asserted or queried.
We reflect this condition in the predicate
\begin{align}
	\tag{is\_rel}
	\label{eq:relevance-condition}
	\begin{split}
	\isrelevant(b, \ctrl) \quad := \qquad
		&b = 1 \qtext{or} \ct \in \ctrl \quad\text{or}\\
		&b = 0 \qtext{or} \cf \in \ctrl.
	\end{split}
\end{align}
It applies to constraints in the form $\cbw Cb{\ctrl}$.
We show later that this principle can be applied to primitive constraints.


\subsection{Boolean Constraints}

We again focus on disjunctive constraints.
The following rule decomposes the constraint
$\cb {(C_1 \lor C_2)}b$
only if the relevance test is passed.
In this case the shared control sets are initialised.
\begin{align}
	\tag{$\cor_{\text{dec}}$}
	\label{rule:or-decomposition}
	\begin{split}
	\begin{array}[b]{@{}l}
	\isrelevant(b, \ctrl)
	\end{array}
	\entails &
		\cwt{\cor(b, b_1, b_2)}{\ctrl, \ctrl_1, \ctrl_2},
		\\&
		\cbw {C_1}{b_1}{\ctrl_1}, \ctrl_1 := \emptyset,
		\\&
		\cbw {C_2}{b_2}{\ctrl_2}, \ctrl_2 := \emptyset.
	\end{split}
\end{align}
The following rules specify how control information propagates
through this disjunctive constraint
in accordance with Proposition~\ref{prop:disj}:
\begin{align}
	b = 1	&\entails
		\ctrl_1 \setadd \cf,
		\ctrl_2 \setadd \cf;
	\nonumber
	\\
	b_1 = 0	&\entails
		\ctrl \trails \ctrl_2,
		\del\ \cor(b, b_1, b_2);
	\nonumber
	\\
	b_2 = 0	&\entails
		\ctrl \trails \ctrl_1,
		\del\ \cor(b, b_1, b_2);
	\nonumber
	\\
	b_1 = 1	&\entails
		\ctrl_2 \setadd \ir,
		\del\ \cor(b, b_1, b_2);
	\nonumber
	\\
	b_2 = 1	&\entails
		\ctrl_1 \setadd \ir,
		\del\ \cor(b, b_1, b_2);
	\nonumber
	\\
	\cf \in \ctrl	&\entails
		\ctrl_1 \setadd \cf
	\tag{$\cor_{\text{cf}}$}
	\label{rule:or-control-checkfalse};
	\\
	\ct \in \ctrl	&\entails
		\ctrl_1 \setadd \ct,
		\ctrl_2 \setadd \ct;
	\nonumber
	\\
	\ir \in \ctrl	&\entails
		\ctrl_1 \setadd \ir,
		\ctrl_2 \setadd \ir,
		\del\ \cor(b, b_1, b_2).
	\nonumber
\end{align}
In rule~\eqref{rule:or-control-checkfalse},
we arbitrarily select the first disjunct to receive $\cf$.
For comparison and completeness, here are the rules
propagating truth information:
\begin{align*}
	b_1 = 0	&\entails	b = b_2;		&
	b_1 = 1	&\entails	b = 1;			\\
	b_2 = 0	&\entails	b = b_1;		&
	b_2 = 1	&\entails	b = 1;			\\
	b   = 0	&\entails	b_1 = 0, b_2 = 0.
\end{align*}

\newcommand{\bnot} {N}
\noindent
Control propagation for the negation constraint
$\cwt{\cnot(b, b_{\bnot})}{\ctrl, \ctrl_{\bnot}}$ is straightforward:
\begin{align*}
	b = 1	\text{ or } b = 0 \text{ or }
	b_{\bnot} = 1	\text{ or } b_{\bnot} = 0
			&\entails	\del\ \cnot(b, b_{\bnot});	\\
	\cf \in \ctrl 	&\entails	\ctrl_{\bnot} \setadd \ct;	\\
	\ct \in \ctrl	&\entails	\ctrl_{\bnot} \setadd \cf;	\\
	\ir \in \ctrl 	&\entails	\ctrl_{\bnot} \setadd \ir.
\end{align*}
The rules for other Boolean operators are analogous.
Note that a move from binary to $n$-ary conjunctions or disjunctions
does not affect the control flow in principle, in the same way that
the logic is unaffected.

Both $\ct$ and $\cf$ can be in the control set of a constraint
at the same time,
as it might be in a both positive and negative context.
An example is the condition of an if-then-else.
On the other hand,
if for instance a constraint is not in a negated context, $\cf$ cannot arise.


\subsection{Primitive Constraints}

Asserting and querying other primitive constraints
can be controlled similarly to Boolean constraints.
In particular, the relevance condition~\eqref{eq:relevance-condition}
must be satisfied before inspecting a constraint.
We furthermore deal with $\ir \in \ctrl$ as expected, by not asserting
the primitive constraint or by deleting it from the set of currently
queried or asserted constraints.

When a query on a primitive constraint is inconclusive,
it is re-evaluated whenever useful.
This can be when elements from a variable domain are removed or when a bound changes.
We rely on the constraint solving environment to signal such changes.

Deciding the truth or the falsity of a constraint in general is
an expensive operation that requires the evaluation of every
variable domain.
A primitive $C(X)$ is guaranteed to be true if and only if
$C(X) \subseteq D(X)$ and $C(X)$ is non-empty.
$C$ is guaranteed to be false if and only if
$C(X) \setintersection D(X) = \emptyset$,
where $X = \rangen x$ and $D(X) = D(x_1) \times \ldots \times D(x_n)$.
For some primitive constraints
we can give complete but simpler evaluation criteria,
similarly to indexicals \cite{codognet:1996:compiling};
see Tab.~\ref{tab:primitive-constraint-queries}.
\begin{table}
\[\begin{array}{@{}l@{\quad}||@{\quad}l|@{\quad}l}
\text{Constraint }
 	& \text{true if}
 	& \text{false if}			\\[1ex]\hline&&\\[-1.5ex]
x \in S
	& D(x) \subseteq S
	& D(x) \setintersection S = \emptyset		\\[0.5ex]
x = a
	& |D(x)| = 1,\ D(x) = \{a\}
	& a \notin D(x) 				\\[0.5ex]
x = y
	& |D(x)| = |D(y)| = 1,\ D(x) = D(y)\quad
	& D(x) \setintersection D(y) = \emptyset	\\[0.5ex]
x \leq y
	& \max(D(x)) \leq \min(D(y))
	& \min(D(x)) > \max(D(y))
\end{array}\]
\caption{Primitive constraint queries
($S$ is a constant set, $a$ is a constant value)}
\label{tab:primitive-constraint-queries}
\end{table}

Practical constraint solving systems usually maintain domain bounds explicitly.
This makes answering the truth query for equality constraints and
the queries for ordering constraints very efficient.
Furthermore, the re-evaluation of a query can be better controlled:
only changes of the respective bounds are an event that makes a
re-evaluation worthwhile.


\section{Implied Constraints}
\label{sec:implied-constraints}

Appropriate handling of implied constraints fits naturally into
the control propagation framework.
Suppose the disjunctive constraint $C_1 \lor C_2$ implies $C_{\impl}$;
that is, $(C_1 \lor C_2) \limplies C_{\impl}$ is always true.
Logically, $C_{\impl}$ is redundant.
In terms of constraint propagation, it may not be, however.

Consider the disjunction $(x = y) \lor (x < y)$, which implies $x \leq y$.
Assume the domains are $D(x) = \{4,5\}$, $D(y) = \{3,4,5\}$.
Since the individual disjuncts are not false, there is
no propagation from the decomposition.
In order to conclude $x \leq y$ and thus $D(y) = \{4,5\}$
we associate the constraint with its implied constraint.

We write a disjunctive constraint annotated with an implied constraint as
\[
	C_1 \lor C_2 \lentailed C_{\impl}.
\]
To benefit from the propagation of $C_{\impl}$, we could
represent this constraint as $(C_1 \lor C_2) \land C_{\impl}$.
However, this representation has the shortcoming that it leads
to redundant propagation in some circumstances.
Once one disjunct, say, $C_1$, is known to be false,
the other disjunct, $C_2$, can be imposed.
The propagation of $C_{\impl}$ is then still executed, however,
while it is subsumed by that of $C_2$.  It is desirable to
recognise that $C_{\impl}$ is operationally redundant at this point.
We capture this situation by enhancing the decomposition
rule~\eqref{rule:or-decomposition} as follows:
\begin{align*}
	\begin{split}
	\cbw {(C_1 \lor C_2 \lentailed C_{\impl})}b{\ctrl} \entails&
		\cwt{\corimpl(b, b_1, b_2, b_{\impl})}%
			{\ctrl, \ctrl_1, \ctrl_2, \ctrl_{\impl}},	\\&
			\cbw {C_1}{b_1}{\ctrl_1}, \ctrl_1 := \emptyset, \\&
			\cbw {C_2}{b_2}{\ctrl_2}, \ctrl_2 := \emptyset, \\&
			\cbw {C_{\impl}}{b_{\impl}}{\ctrl_{\impl}}, \ctrl_{\impl} := \emptyset.
	\end{split}
\end{align*}
Additionally to the control rules for regular disjunctive constraints
shown earlier, we now also use the following four rules:
\newcommand{\xentails}{\ \limpliesd\ }
\begin{align*}
	\!\!b_{\impl} = 0	&\xentails	b = 0;
		\hspace{1.8em} &
	b_1 = 0	&\xentails	\ctrl_{\impl} \setadd \ir,
				\del\ \corimpl(b, b_1, b_2, b_{\impl});\\
	b = 1	&\xentails		b_{\impl} = 1;	&
	b_2 = 0 &\xentails	\ctrl_{\impl} \setadd \ir,
				\del\ \corimpl(b, b_1, b_2, b_{\impl}).
\end{align*}

\noindent
We envisage the automated discovery of implied constraints,
but for now we assume manual annotation.


\section{Subconstraint Sharing: From Trees to DAGs}

The straightforward decomposition of complex constraints
can contain unnecessary copies of the same subconstraint
in different contexts.  The dual constraint graph (whose vertices
are the constraints and whose edges are the variables)
is a tree, while often a directed acyclic graph
(DAG) gives a logically equivalent but more compact representation.
See, for example, CDDs \cite{cheng:2005:constrained}.

\begin{figure}
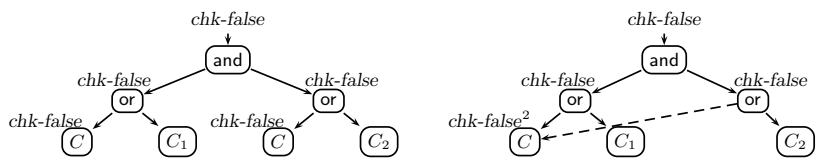

\center
\scalebox{0.8}{\psset{unit=1ex}
\psset{treesep=8,levelsep=5}
\pstree[nodesepA=1]{\Tr{\cf}}{
	\pstree[nodesepA=0]{\Trf{\gand}}{
		\pstree{\Trf{\gor}\tlput{$\cf$}}{
			\Trf{$C$}\tlput{$\cf$}
			\Trf{$C_1$}
		}
		\pstree{\Trf{\gor}\trput{$\cf$}}{
			\Trf{$C$}\tlput{$\cf$}
			\Trf{$C_2$}
		}
	}
}
\hspace{5em}
\pstree[nodesepA=1]{\Tr{\cf}}{
	\pstree[nodesepA=0]{\Trf{\gand}}{
		\pstree{\Trf{\gor}\tlput{$\cf$}}{
			\Tr[name=c1left]{\psframebox{$C$}}\tlput{$\cf^2$}
			\Trf{$C_1$}
		}
		\pstree{\Tr[name=orright]{\psframebox{\gor}}\trput{$\cf$}}{
			\Tn
			\Trf{$C_2$}
		}
	}
}
\ncline[linestyle=dashed]{<-}{c1left}{orright}
}
\caption{Left: no sharing.  Right: sharing with reference counting}
\label{fig:subconstraint-sharing}
\end{figure}

We can apply controlled propagation to complex constraints
represented in DAG form.
We need to account for the multiplicity of a constraint when
handling queries on it:
the set of control flags now becomes a \emph{multiset}, and
in effect, we maintain \emph{reference counters} for subconstraints.
Control flags need to be properly subtracted from the control
set of a constraint.
For the sake of a simple example, consider the constraint
$(C \lor C_1) \land (C \lor C_2)$.
Fig.~\ref{fig:subconstraint-sharing} shows a decomposition of it.

Another example is the condition in an if-then-else constraint.
Opportunities for shared structures arise frequently when constraints
are defined in terms of subconstraints that in turn are constructed
by recursive definitions.


\section{Case Studies}
\label{sec:case-studies}

We examine several constraints studied in the literature
and show that their decomposition benefits from controlled propagation.

\medskip
\noindent
{\bf Literal Watching.}
The DPLL procedure for solving the SAT problem
uses a combination of search and inference
and can be viewed as a special case of constraint programming.
Many SAT solvers based on DPLL employ unit propagation with
\emph{2-literal watching}, \eg\ Chaff \cite{moskewicz:2001:chaff}.
At any time, only changes to two literals per clause are tracked,
and consideration of other literals is postponed.

\begin{figure}
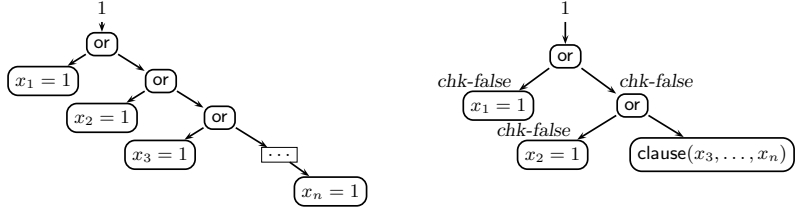

\center\scalebox{0.8}{\psset{unit=1ex}
\psset{treesep=8,levelsep=4.5}
\pstree[nodesepA=4pt]{\Tr{\colorlogic 1}}{
	\pstree{\Trf{\gor}}{\psset{nodesepA=0pt}\Trf{$x_1 = 1$}
	\pstree{\Trf{\gor}}{\Trf{$x_2 = 1$}
	\pstree{\Trf{\gor}}{\Trf{$x_3 = 1$}
	\pstree{\Tr{\psframebox[linewidth=0,framearc=0]{$\ldots$}}}{
 		\Tn\Trf{$x_n = 1$}}
}}}}
\hspace{10ex}
\psset{treesep=5,levelsep=6}
\pstree[nodesepA=4pt]{\Tr{\colorlogic 1}}{
	\pstree{\Trf{\gor}}{\psset{nodesepA=0pt}\Trf{$x_1 = 1$}\tlput{$\cf$}
	\Tn\pstree{\Trf{\gor}\trput{$\cf$}}{\Trf{$x_2 = 1$}\tlput{$\cf$}
	\Trf{$\clause(\rangeb x3n)$}
}}}
}
\caption{Uncontrolled versus controlled decomposition of $\clause$}
\label{fig:clause-constraint}
\end{figure}

Let us view a propositional clause as a Boolean constraint.
We define
\[
	\clause(\rangen x) \quad := \qquad
		x_1 = 1 \ \lor\  \clause(\rangeb x2n)
\]
and show in Fig.~\ref{fig:clause-constraint}
the decomposition of $\clause(\rangen x)$ as a graph
for controlled and uncontrolled propagation
(where $D(x_i) = \{0,1\}$ for all $x_i$).
Both propagation variants enforce domain-consistency
if the primitive equality constraints do and the variables are pairwise different.
This corresponds to unit propagation.

Uncontrolled decomposition expands fully into
$n-1$ Boolean $\cor$ constraints and $n$ primitive constraints $x_i = 1$.
Controlled decomposition
only expands into two $\cor$ constraints and
the first two primitive constraints $x_1=1$, $x_2=1$.
The leaf node marked $\clause(\rangeb x3n)$
is initially not expanded as neither assertion
nor query information is passed to it.
The essence is that the first $\cor$ constraint results in two
$\cf$ queries to the subordinate $\cor$ constraint which passes
this query on to just one disjunct.
This structure is maintained with respect to new information
such as variable instantiations.
No more than two primitive equality constraints are ever queried at a time.
A reduction of inference effort as well as of space usage results.

Controlled propagation here corresponds precisely to 2-literal watching.


\medskip
\noindent
{\bf Disequality of Tuples.}
Finite domain constraint programming generally focuses on variables
over the integers.
Sometimes, higher-structured variable types, such as sets of integers,
are more appropriate for modelling.
Many complex constraints studied in the constraint community
are on a sequence of variables and can thus naturally be
viewed as constraining a variable whose type is tuple-of-integers.
The recent study \cite{quimper:2005:beyond}
examines how some known constraint propagation algorithms
for integer variables can be lifted to higher-structured variables.
One of the constraints examined is $\constraint{alldifferent}$ on tuples,
which requires a sequence of variables of type tuple-of-integers
to be pairwise different.
Its straightforward definition is
\[
	\alldifft(\tuple{\rangen X}) \ := \
		\lAnd_{i,j \in 1, \ldots, n,\ i < j}
		\diff(X_i, X_j),
\]
where
\[
	\diff(\tuple{\rangeb x1m}, \tuple{\rangeb y1m}) \ := \
		\lOr_{i \in 1, \ldots, m} x_i \neq y_i.
\]

\noindent
Let us examine these constraints with respect to controlled propagation.
The $\diff$ constraint is a large disjunction, and it
behaves thus like the $\clause$ constraint studied in the previous
section -- at most two disjuncts $x_i \neq y_i$ are queried
for falsity at any time.

Deciding the falsity of a disequality constraint
is particularly efficient when the primitive constraints in
Tab.~\ref{tab:primitive-constraint-queries} are used, \ie\
falsity of disequality when the domains are singletons.
If the domains are not singletons,
re-evaluation of the query is only necessary once that is the case.
In contrast, a truth query for a disequality is
(more) expensive as the domains must be intersected, and,
if inconclusive, should be re-evaluated whenever any domain change occurred.

The $\alldifft$ constraint is a conjunction of $\binom n2$
$\diff$ constraints.
Therefore, controlled propagation queries
at most $n(n-1)$ disequality constraints for falsity at a time.
Uncontrolled propagation asserts all $n(n-1)m/2$ reified disequality constraints
and in essence queries truth and falsity of each.
Using controlled rather than uncontrolled decomposition-based
propagation for $\alldifft$ saves substantial effort
without loss of effective propagation.

We remark that a specialised, stronger but non-trivial propagation algorithm
for this case has been studied in \cite{quimper:2005:beyond}.
The controlled propagation framework is then useful when
specialised algorithms are not readily available, for example due to
a lack of expertise or resources in the design and implementation
of propagation algorithms.


\medskip
\noindent
{\bf Lexicographic Ordering Constraint.}
It is often desirable to prevent symmetries in constraint problems.
One way is to add symmetry-breaking constraints such as the
lexicographic ordering constraint \cite{frisch:2002:global}.
A straightforward definition is as follows:
\[
	\lex(\tuple{\rangen x}, \tuple{\rangen y}) \ := \
	\begin{array}[t]{@{}l}
		x_1 < y_1 \\
		\lor\\
		x_1 = y_1 \land \lex(\tuple{\rangeb x2n}, \tuple{\rangeb y2n})\\
		\lor\\
		n = 0
	\end{array}
\]
With this definition, propagation of the decomposition does not
always enforce domain-consistency.  Consider
$\lex(\tuple{x_1, x_2}, \tuple{y_1, y_2})$ with the domains
$D(x_1) = D(x_2) = D(y_2) = \{3..5\}$ and $D(y_1) = \{0..5\}$.
Controlled decomposition results in the reified versions of
$x_1 < y_1$, $x_1 = y_1$, $x_2 < y_2$
connected by Boolean constraints.
None of these primitive constraints is true or false.
Yet we should be able to conclude $x_1 \leq y_1$, hence $D(y_1) = \{3..5\}$,
from the definition of $\lex$.

The difficulty is that the naive decomposition is weaker than the logical
definition because it only reasons on the individual primitive constraints.
However, it is easy to see that $x_1 \leq y_1$ is an implied constraint
in the sense of Section~\ref{sec:implied-constraints},
and we can annotate the definition of $\lex$ accordingly:
\[
	\lex(\tuple{\rangen x}, \tuple{\rangen y}) \ := \quad
	\begin{array}[t]{@{}l}
		x_1 < y_1 \\
		\lor\\
		x_1 = y_1 \land \lex(\tuple{\rangeb x2n}, \tuple{\rangeb y2n})\\
		\lentailed x_1 \leq y_1\\
		\lor\\
		n = 0
	\end{array}
\]
We state without proof that propagation of the constraints of the decomposition
enforces domain-consistency on $\lex$ if the annotated definition is used.

Tab.~\ref{tab:lex-derivation} represents a trace of
$\lex$ on the example used in \cite{frisch:2002:global},
showing the lazy decomposing due to controlled propagation.
We collapse several atomic inference steps and omit the Boolean constraints,
and we write $v_{i..j}$ to abbreviate $\rangeb vij$.
Observe how the implied constraints $x_i \leq y_i$ are asserted,
made irrelevant and then deleted.
The derivation ends with no constraints other than $x_3 < y_3$ queried or asserted.

\begin{table}
\center
\scalebox{0.92}{
$\begin{array}{@{\extracolsep{1ex}}ll|ll|llllll}
\text{Asserted} &&
\text{Set of constraints queried for} &&
\multicolumn 3l{\text{Variable domains}}\\
&&\text{falsity}&&& x_1 & x_2 & x_3 & x_4 & x_5\\
&&&&& y_1 & y_2 & y_3 & x_4 & y_5\\\hline
\lex(\tuple{x_{1..5}}, &&&& \\
\hspace*{2em} \tuple{y_{1..5}})&&&&
\\&&&&
&\{2\} & \{1,3,4\} & \{1..5\} & \{1..2\} & \{3..5\}\\&&&&
&\{0..2\} & \{1\} & \{0..4\} & \{0..1\} & \{0..2\}
\\
x_1 \leq y_1 &&		x_1 < y_1, x_1 = y_1, x_2 < y_2 &&
\\
&&&&
&\{2\} & \{1,3,4\} & \{1..5\} & \{1..2\} & \{3..5\}\\&&&&
&\mathbf{\{2\}} & \{1\} & \{0..4\} & \{0..1\} & \{0..2\}
\\
x_2 \leq y_2 &&		x_2 < y_2, x_2 = y_2, x_3 < y_3 &&
\\&&&&
&\{2\} & \mathbf{\{1\}} & \{1..5\} & \{1..2\} & \{3..5\}\\&&&&
&\{2\} & \{1\} & \{0..4\} & \{0..1\} & \{0..2\}
\\
x_3 \leq y_3 &&		x_3 < y_3, x_3 = y_3, x_4 < y_4 &&
\\&&&&
&\{2\} & \{1\} & \mathbf{\{1..4\}} & \{1..2\} & \{3..5\}\\&&&&
&\{2\} & \{1\} & \mathbf{\{1..4\}} & \{0..1\} & \{0..2\}
\\
x_3 \leq y_3 &&		x_3 < y_3, x_3 = y_3, x_4 = y_4, x_5 < y_5 &&
\\&&&&
&\{2\} & \{1\} & \{1..4\} & \{1..2\} & \{3..5\}\\&&&&
&\{2\} & \{1\} & \{1..4\} & \{0..1\} & \{0..2\}
\\
x_3 \leq y_3 &&		x_3 < y_3, x_3 = y_3, x_4 = y_4, x_5 = y_5 &&
\\&&&&
&\{2\} & \{1\} & \{1..4\} & \{1..2\} & \{3..5\}\\&&&&
&\{2\} & \{1\} & \{1..4\} & \{0..1\} & \{0..2\}
\\
x_3 < y_3 &&
&&
\\&&&&
&\{2\} & \{1\} & \mathbf{\{1..3\}} & \{1..2\} & \{3..5\}\\&&&&
&\{2\} & \{1\} & \mathbf{\{2..4\}} & \{0..1\} & \{0..2\}
\end{array}$
}
\vspace{2ex}
\caption{An example of controlled propagation of the $\lex$ constraint}
\label{tab:lex-derivation}
\vspace{-4ex}
\end{table}


\section{Implementation and Benchmarks}

We implemented a prototype of the controlled propagation framework
in the CLP system \eclipse{} \cite{wallace:1997:eclipse},
using its predicate suspension features
and attributed variables to handle control information.
The implementation
provides controlled propagation for the basic Boolean
and primitive constraints, and it handles implied constraints.
Structure-sharing by a DAG-structured decomposition is not supported.

We conducted several simple benchmarks to compare controlled
and uncontrolled propagation on constraint decompositions,
using the $\clause$, $\diff$, $\alldifft$ and $\lex$ constraints.
A benchmark consisted of finding a solution to a single constraint.
For the uncontrolled propagation benchmark,
the constraint was simply decomposed
into built-in Boolean and primitive constraints of \eclipse,
and implied constraints (in $\lex$) were conjunctively
added to their respective premise.

The number of variables in the respective tuple(s) was varied between five and 50.
For the $\alldifft$ benchmark, we chose $20$ tuples.
The variables ranged over the interval $\{1..10\}$ (except for $\clause$).
Solutions to the constraints were searched
by randomly selecting a variable and a value in its domain.
This value was either assigned or excluded from its domain; this choice was also random.
To obtain meaningful averages, every individual solution search was run a sufficient
number of times (typically a few 10000)
so that the total computation time was roughly 15\;s.
Each of these runs used a new initialisation
of the pseudo-random number generator resulting in a possibly different solution,
while the benchmark versions (controlled vs.\ uncontrolled propagation)
used the same initial value to obtain identical search trees.
Every experiment was repeated five times.
In Tab.~\ref{tab:benchmarks},
we give the relative solving time with controlled propagation,
based on the corresponding uncontrolled propagation benchmark taken to be 100\%.
\begin{table}[t]
\center\small
\begin{tabular}{@{\extracolsep{0.5em}}|@{\hspace{0.5em}}l||*4{rrrr|}}
	\hline
	&
	\multicolumn 4l{$\clause$}	\vline&
	\multicolumn 4l{$\diff$}	\vline&
	\multicolumn 4l{$\alldifft$}	\vline&
	\multicolumn 4l{$\lex$}		\vline
	\\[1ex]
	nb.\ of variables
	& 5	& 10	& 20	& 50
	& 5	& 10	& 20	& 50
	& 5	& 10	& 20	& 50
	& 5	& 10	& 20	& 50
	\\\hline\hline
	runtime (\%)
	& 100 & 69 & 50 & 38	
	&  88 & 84 & 67 & 62	
	&  66 & 38 & 23 & 11	
	& 138 & 92 & 69 & 54	
	\\\hline
\end{tabular}
\vspace{1ex}
\caption{Benchmark results: controlled propagation (uncontrolled prop.\ = 100\%)}
\label{tab:benchmarks}
\vspace{-4ex}
\end{table}

The benchmarks show that controlling propagation can reduce the propagation time.
The reduction is especially substantial for high-arity constraints.
For low-arity constraints, the extra cost of maintaining control information
in our implementation can outweigh the saving due to less propagation.
While we have not measured the space usage of the two propagation approaches,
it follows from the analyses in Section~\ref{sec:case-studies}
that using controlled propagation for the considered constraints
often also requires less space,
since constraints are decomposed only when required.

We remark that efficiency was a minor concern in our high-level,
proof-of-concept implementation;
consequently we expect that it can be improved considerably.
For example, for constraints that are in negation normal form
(all constraints in our benchmark), the control flag $\ct$ is never created.
A simpler subset of the control propagation rules can then be used.


\section{Final Remarks}

\subsubsection*{Related Work.}

In terms of foundations, the controlled propagation framework can be described
as a refined instance of the CLP scheme (see \cite{jaffar:1994:constraint}),
by a subdivision of the set of active constraints according to their associated
truth and falsity queries.
Concurrent constraint programming (CCP) \cite{saraswat:1993:ccp},
based on asserting and querying constraints, is closely related;
our propagation framework can be viewed as an extension in which
control is explicitly addressed and dealt with in a fine-grained way.
A practical CCP-based language such as CHR \cite{fruehwirth:1998:theory}
would lend itself well to an implementation.
For example, control propagation rules
with $\del$ statements can be implemented as simplification rules.

A number of approaches address the issue of propagation of complex constraints.
The proposal of \cite{bacchus:2005:propagating}
is to view a constraint as an expression from which sets of inconsistent
or valid variable assignments (in extension) can be computed.
It focuses more on the complexity issues of achieving certain kinds
of local consistencies.
The work \cite{beldiceanu:2004:deriving} studies
semi-automatic construction of propagation mechanisms
for constraints defined by extended finite automata.
An automaton is captured by signature (automaton input) constraints
and state transition constraints.
Signature constraints represent groups of reified primitive constraints
and are considered pre-defined.
They communicate with state transition constraints via constrained variables,
which correspond to tuples of Boolean variables
of the reified constraints in the signature constraints.
Similarly to propagating the constraint in decomposition,
all automata constraints are propagated independently of each other.

Controlled propagation is similar to techniques used in \textsc{NoClause},
a SAT solver for propositional non-CNF formulas \cite{thiffault:2004:solving},
which in turn lifts techniques such as 2-literal watching from CNF to non-CNF
solvers.  We describe here these techniques in a formal, abstract framework and
integrate non-Boolean primitive constraints and implied constraints,
thus making them usable for constraint propagation.


\subsubsection*{Conclusion.}

We have proposed a new framework for propagating arbitrary complex constraints.
It is characterised by viewing logic and control as separate concerns.
We have shown that the controlled propagation framework
explains and generalises some of the principles on which
efficient manually devised propagation algorithms
for complex constraints are based.
By discussing an implementation and benchmarks,
we have demonstrated feasibility and efficiency.
The practical benefits of the controlled propagation framework are
that it provides \emph{automatic} constraint propagation for
\emph{arbitrary} logical combinations of primitive constraints.
Depending on the constraint, controlling the propagation
can result in substantially reduced usage of time as well as space.

Our focus in this paper has been on reducing unnecessary inference steps.
The complementary task of automatically identifying and enabling useful
inference steps in our framework deserves to be addressed.
It would be interesting to investigate if automatic reasoning
methods can be used to strengthen constraint definitions,
for instance by automatically deriving implied constraints.


\section*{Acknowledgements}

We thank the anonymous reviewers for their comments.
This paper was written while Roland Yap was visiting
the Swedish Institute of Computer Science and their support
and hospitality are gratefully acknowledged.
The research here is supported by a NUS ARF grant.


\bibliographystyle{alpha}

\newcommand{\etalchar}[1]{$^{#1}$}

\end{document}